# Ultra-sensitive hybrid diamond nanothermometer

Chu-Feng Liu*[1], Weng-Hang Leong*[1], Kangwei Xia*[1], Xi Feng[1], Amit Finkler[§3], Andrej Denisenko[3], Jörg Wrachtrup[3,4], Quan Li[1,2], and Ren-Bao Liu[1,2]

1. Department of physics, The Chinese University of Hong Kong, Shatin, New Territories, Hong Kong, China.

2. The Hong Kong Institute of Quantum information Science and Technology, The Chinese University of Hong Kong, Shatin, New Territories, Hong Kong, China.

3. 3rd Institute of Physics and Center for Applied Quantum Technologies, University of Stuttgart, 70569 Stuttgart, Germany.

4. Max Planck Institute for Solid State Research, 70569 Stuttgart, Germany.

* These authors contributed equally: Chu-Feng Liu, Weng-Hang Leong, Kangwei Xia. Correspondence and request for materials should be addressed to Q.L. (email: liquan@phy.cuhk.edu.hk) or to R.-B.L. (email: rbliu@cuhk.edu.hk).

§ Present affiliation: Department of Chemical and Biological Physics, Weizmann Institute of Science, Rehovot 7610001, Israel

**Abstract**

Nitrogen-vacancy (NV) centers in diamond are promising quantum sensors for their long spin coherence time under ambient conditions. However, their spin resonances are relatively insensitive to non-magnetic parameters such as temperature. A magnetic-nanoparticle-nanodiamond hybrid thermometer, where the temperature change is converted to the magnetic field variation near the Curie temperature, was demonstrated to have enhanced temperature sensitivity (11 mK Hz$^{-1/2}$) [Phys. Rev. X 8, 011042 (2018)], but the sensitivity was limited by the large spectral broadening of ensemble spins in nanodiamonds. To overcome this limitation, here we showed an improved design of a hybrid nanothermometer using a single NV center in a diamond nanopillar coupled with a single magnetic nanoparticle of copper-nickel alloy, and demonstrated a temperature sensitivity of 76 μK Hz$^{-1/2}$. This hybrid design enabled detection of 2 millikelvins temperature changes with temporal resolution of 5 milliseconds. The ultra-sensitive nanothermometer offers a new tool to investigate thermal processes in nanoscale systems.

**Introduction**

Nanoscale temperature measurement with high sensitivity is important to investigating many phenomena such as thermal mapping of nano-/micro-electronics [1], thermoplasmonics of nanoparticles [2], chemical reactions in nanoliter volume [3], and thermal processes in live systems [4–6]. To probe the thermal dynamics on the nanoscale, various measurement protocols have been developed. The optical thermometers convert the local temperature variance to changes of optical lifetime [7], fluorescence intensity [8], Raman shift [9], or emission spectrum [10]. Being a non-contact and convenient method, optical-based nanothermometers, like fluorescence proteins [11], dyes [7], and rare-earth nanoparticles [12], have been proposed and demonstrated for temperature detection under various conditions. However, this method has a relatively low sensitivity (typically $1 \text{ K Hz}^{-1/2}$) [13–15] and moreover some optical sensors are subject to artifacts induced by the local environments of the sensors such as reflective index and pH values [16]. Electronic temperature measurements, such as the scanning thermal microscopy and the superconducting quantum interference device (SQUID), have high spatial resolution and high sensitivity ($\sim 1 \text{ μK Hz}^{-1/2}$) [17,18], but they require extreme operating conditions and are subjected to contact-related artifacts [19,20].

The recent development of a diamond-based thermometer poses a promising alternative [21–23]. Nitrogen-vacancy (NV) centers in diamond have long spin coherence time under ambient conditions [24]. Their spin resonance frequencies shift with the environmental temperature [25], which is robust against artifacts from the local environment. With photo-stability of NV centers [26], high thermal conductivity [27] and bio-compatibility of diamond material [28,29], diamond-based thermometers are a potential candidate for temperature sensing in complex systems without the requirement of extreme operating conditions. However, the temperature dependence of NV center spin transition frequencies ($dD/dT \approx -74 \text{ kHz K}^{-1}$) is relatively small. Thus there arises the idea of hybrid diamond thermometers [30,31], in which the temperature change is transduced to a magnetic signal to be detected by the NV center spin resonance. A hybrid nanothermometer composed of a single copper-nickel alloy magnetic nanoparticle (MNP) and a diamond nanocrystal with ensemble NV centers [30] was demonstrated to have a sensitivity as high as $11 \text{ mK Hz}^{-1/2}$ near the Curie temperature of the magnetic nanoparticle, where a small temperature change leads to a large magnetic field change due to the critical magnetization. However, the sensitivity of this hybrid nanothermometer was limited by the short coherent time of ensemble NV centers in nanodiamonds as well as the ODMR linewidth broadening due to the large magnetic field gradient of from the magnetic nanoparticle. To overcome this limitation, we constructed a hybrid nanothermometer employing a single NV center in a diamond nanopillar and a single copper-nickel alloy nanoparticle. This design has the following advantages: the spin coherence time of the single NV center is much longer than that in nanodiamonds, and the field gradient induced broadening for ensemble NV centers in nanodiamond is eliminated [30]. Although the photon count rate of a single NV is lower than that of ensembles in nanodiamond, the pillar waveguide configuration largely enhance the photon collection efficiency [32]. We constructed the hybrid nanothermometer by placing the magnetic nanoparticle close to the diamond nanopillar via atomic force microscopy (AFM) nanomanipulation. Such a hybrid nanothermometer has a temperature sensitivity of $76 \text{ μK Hz}^{-1/2}$. To the best of our knowledge, this is the most sensitive nanothermometer working under ambient conditions. By employing this hybrid sensor, we monitor the temperature changes of a laser heating process and environment temperature fluctuations, as well as thermal dissipation near the sensor when additional heating to the system is induced by controlling the current passing through the microwave antenna. This ultra-sensitive hybrid nanothermometer offers the opportunities of studying fast thermal processes in nanostructures and/or in living systems.

**Results**

The hybrid diamond nanothermometer is composed of a single NV center in a diamond pillar and a copper-nickel alloy MNP, as illustrated in the inset of Fig. 1a. The ground state of an NV center is a spin triplet. The simplified spin Hamiltonian can be written as

$$H = D\mathbf{S}^2 + \gamma_e \mathbf{B} \cdot \mathbf{S}, \tag{1}$$

where $\mathbf{S}$ is the spin operator, $\mathbf{B}$ is the external magnetic field, $D \approx 2.87$ GHz is the zero-field splitting between the $m_s = 0$ and the $m_s = \pm 1$ states, and the electron gyromagnetic ratio $\gamma_e = 2.8$ MHz Gauss$^{-1}$. The transition frequencies between different spin states can be measured by optically detected magnetic resonance (ODMR) spectroscopy using the spin-dependent fluorescence and resonant microwave manipulation of the spin. Unlike conventional diamond thermometry based on the temperature dependence of $D$, which has a susceptibility of $dD/dT \approx -74$ kHz K$^{-1}$, the hybrid nanothermometer measures the magnetization change of the MNP as induced by the temperature variation. Near the critical point of the MNP, the temperature susceptibility is large and hence a high temperature sensitivity can be achieved. Figure 1a shows a simulated demagnetization curve when an MNP undergoes the ferromagnetic-paramagnetic transition under a small external magnetic field (100 Gauss). The magnetization of the MNP changes drastically when the temperature approaches the Curie point ($T_C$). The magnetic field from the MNP induces the Zeeman splitting between the $m_s = -1$ and $m_s = +1$ states of the NV center in diamond, which can be measured through the ODMR spectroscopy. Using a single NV center in a diamond nanopillar has several advantages over the previous hybrid configuration [30]. First, single NV centers in diamond have longer coherence times than ensemble NV centers in nanodiamonds. In our experiments, the selected NV center in nanopillar shows a dephasing time of $T_2^* \sim 1.5$ μs (see Supplementary Figure 2 for details of optical and spin properties of the NV center). As a comparison, the typical dephasing time of NV centers in nanodiamonds is ~100 ns [33]. Second, although the fluorescence intensity of a single NV center is lower than that of ensembles in nanodiamond, the diamond nanopillar acts as a waveguide making the emission more directional and therefore enhancing the fluorescence collection efficiency [32]. Third, the large inhomogeneous broadening of the ODMR of ensemble NV centers in nanodiamond due to the magnetic gradient near the MNP is eliminated in the case of single NV centers.

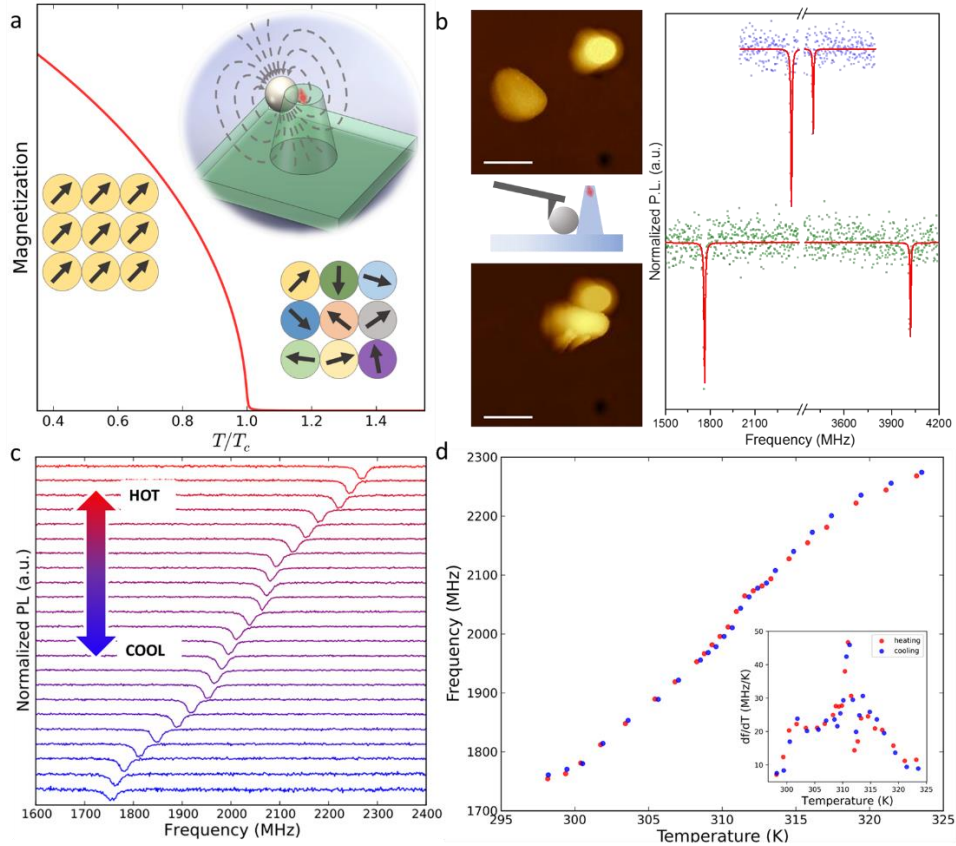

**Figure 1.** Design of a hybrid nanothermometer composed of a single magnetic copper-nickel alloy nanoparticle and a single nitrogen-vacancy (NV) center in a diamond nanopillar. **a**, Simulation of the magnetization **M** of a copper-nickel alloy nanoparticle as a function of temperature under a magnetic field of 100 Gauss. The inset illustrates the configuration of the hybrid nanothermometer. **b**, Atomic force microscopy (AFM) image of the copper-nickel alloy magnetic nanoparticle (MNP) and the diamond nanopillar before the nanomanipulation (upper graph) and after the nanomanipulation (lower graph), and the corresponding optically detected magnetic resonance (ODMR) spectra of the single NV center before and after nanomanipulation (dots being measurement data and lines the double Lorentzian peak fitting). Scale bar is 1 μm. **c**, ODMR spectra of the hybrid nanothermometer at different environment temperatures (from 298 K to 324 K from bottom to top). **d**, ODMR frequency shifts in the heating (blue) and cooling (red) processes. The inset shows the temperature susceptibility of the hybrid nanothermometer, which has the maximum $df/dT \sim 47$ MHz K$^{-1}$ (at a temperature of 311 K).

The hybrid nanothermometer was constructed by nanomanipulation in an AFM setup. The key principle of operation of the hybrid nanothermometer is the effective coupling between the MNP and the NV center, which strongly depends on their separation distance and relative orientation. Figure 1b shows an example of such manipulation. The left panels in Figure 1b show the AFM topographic images of the diamond pillar and a nearby MNP before and after nanomanipulation of the MNP using the AFM tip. A much larger splitting of the $m_s = \pm 1$ states appears in the ODMR spectra when the MNP is pushed closer to the diamond nanopillar, reducing its distance from the NV in the pillar (Fig. 1b right panel). Apart from the coupling strength between the MNP and the NV center, the working range of the hybrid sensor is tens of Kelvin below the Curie temperature. The Curie temperature can be designed by tuning the chemical composition of the copper-nickel alloy nanoparticle. Thus, our hybrid sensor can have a broad working range from cryogenic temperatures to about 600 K [30].

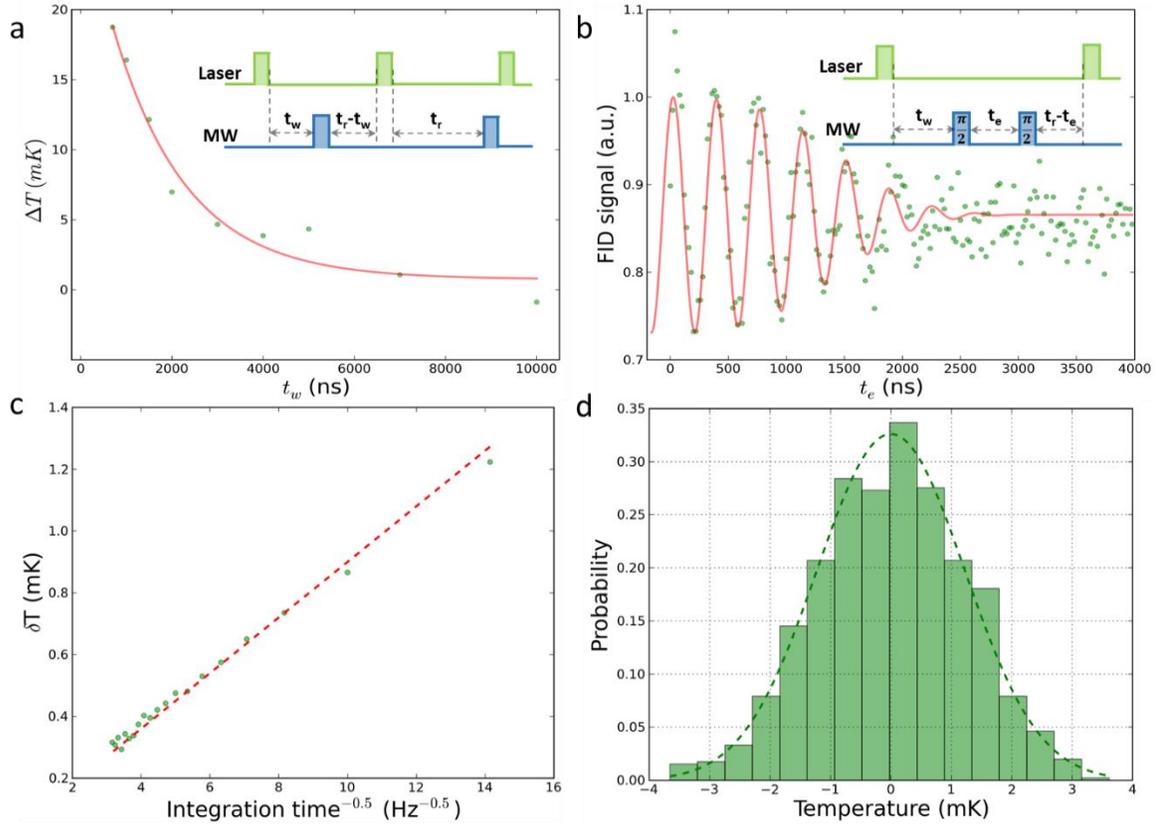

**Figure 2.** Sensitivity of the hybrid nanothermometer. **a**, Cooling curve of the copper-nickel alloy MNP after the laser was turned off. Inset: pulse sequence for measuring the cooling dynamics of the MNP in the hybrid nanothermometer. **b**, Free induction decay (FID) of the NV center spin in the hybrid nanothermometer. The inset shows the pulse sequence. The delay time $t_w$ between the initialization laser and the microwave pulse sequence was chosen to be the cooling time of the MNP ($t_w$ = 1500 ns). **c**, Dependence of the temperature standard deviation on data integration time using FID real-time measurement. the shot-noise-limited sensitivity is derived from the slope of the fitting curve (red dashed line). **d**, A typical histogram of temperature measured in a period of 30 seconds (with sampling time 5 ms).

To characterize the temperature response of the magnetization of the copper-nickel alloy MNP, we measured the magnetic field at the NV center using continuous-wave ODMR spectroscopy. The environment temperature was controlled by a ceramic heater and calibrated by monitoring the $D$ shift of a reference NV center that is far from any MNP (therefore under zero magnetic field) (see Supplementary Figure 3). After the temperature calibration, a magnetic field of 192 Gauss was applied to enhance the local magnetic field generated by the MNP. Figure 1c plots a series of ODMR spectra of the hybrid nanothermometer under different temperatures. The resonance dips indicate the transition between the $m_s = 0$ and $m_s = -1$ states of the NV center spin. The spin resonance frequencies at different temperatures are obtained and plotted in Fig. 1d. By increasing the environment temperature, the resonance frequency splitting is reduced due to the thermal demagnetization of the MNP. The magnetization of the MNP presented a sensitive response. The inset in Fig. 1d summarizes the temperature susceptibility $df/dT$ of the NV center spin resonance frequency ($m_s = -1$ state). At 38 °C, the susceptibility reached its maximum of 47 MHz K$^{-1}$. Comparing to the temperature dependent $D$ shift of an NV center spin, $dD/dT \approx -74$ kHz K$^{-1}$, the temperature susceptibility of the hybrid nanothermometer is enhanced approximately 600-fold.

Furthermore, the magnetization and demagnetization of this MNP is reversible under the external magnetic field (192 Gauss) alignment, as evidenced by the overlap between the temperature responses during the heating and cooling processes (Fig. 1d). The reversibility and chemical stability of the hybrid nanothermometer were further verified by repeating more heating/cooling measurements on the same hybrid sensor at different times (See Supplementary Figure 4).

For high-precision temperature measurement, it is important to exclude the laser heating effect. In conventional optical-based nanothermometers, laser heating on the thermometers induces a local temperature increase [34] and then the measurement of environmental temperature is complicated by the laser heating effect. Laser heating also exists in our hybrid nanothermometer. However, the pulsed ODMR protocol allows to largely reduce the laser heating effect – the laser can be turned off during the spin evolution period in pulsed measurement. The protocol of the pulsed measurement is as follows (see inset of Fig. 2a): First a laser pulse is applied to initialize the spin; then after a waiting time $t_\mathrm{w}$ a microwave is applied; and finally a laser pulse is applied after time $t_\mathrm{r} - t_\mathrm{w}$ to read out the spin state (which also serves to initialize the spin for the next measurement iteration). The interval ($t_\mathrm{r}$) between laser pulses is kept constant to keep the heating effect the same for various waiting time ($t_\mathrm{w}$) taken between the laser and microwave pulses. To understand the cooling dynamics in the hybrid nanothermometer, we carried out pulsed ODMR measurement with different waiting times $t_\mathrm{w}$. The environment temperature was set at 38 °C where the temperature susceptibility attains the maximum $df/dT = 47$ MHz K$^{-1}$. Considering the environment temperature would change with millikelvin scale during the long term measurement, a reference ODMR measurement with waiting time $t_\mathrm{w} = 10$ µs between the laser and microwave pulses was performed simultaneously to calibrate the spin resonance frequency drift due to long term temperature fluctuation. Figure 2a shows the temperature dynamics of the hybrid sensor as a function of the waiting time $t_\mathrm{w}$. It shows that the pulsed laser excitation (300 ns) of the hybrid nanothermometer induced a local temperature increase of about 20 mK. Such an increase of temperature is several times larger than temperature fluctuations of interest in, e.g., nanoelectronics and biological systems [4,17]. After the laser is turned off, the local temperature decays exponentially and recovers to the environment temperature within a time scale of ~1.5 µs. The laser heating effect can be largely reduced by choosing a waiting time $t_\mathrm{w} \geq 1.5$ µs. In the following experiments, we chose $t_\mathrm{w} = 1.5$ µs to reduce the laser heating effect while still having a reasonable measurement duty ratio.

To determine the temperature sensitivity of the hybrid nanothermometer, free-induction decay (FID) of the NV electron spin was measured. The result is plotted in Fig. 2b. The pulse sequence of the FID measurement (inset of Fig 2b) was modified to reduce the laser heating effect ($t_\mathrm{w} = 1.5$ µs) while the interval between the laser pulses was kept constant so that the total laser power applied to the sample was the same for different FID timing (see Supplementary Figure 5 for comparison of FID signal with and without the pulse modification). At the maximum temperature susceptibility point (38 °C), the sensitivity of the hybrid nanothermometer is estimated to be 76 µK Hz$^{-1/2}$ (see Methods for details of the sensitivity estimation). Consistent results are obtained from two other hybrid sensors, revealing the robustness and reproducibility of our hybrid quantum thermometer design (see Supplementary Figure 6). To further verify that the sensitivity was shot-noise limited, we carried out real-time FID measurement with an optimized waiting time of 983 ns (where we have the maximum resonance frequency susceptibility of the FID signal, see Methods for details). The linear dependence of the temperature accuracy (defined as the standard derivation of the temperature measurements $\sigma_T$) on the inverse square root of integration time (see Fig. 2c) indicates the sensitivity was shot-noise limited; with a shot-noise limited sensitivity in the real-time measurement of about 87 μK Hz$^{-1/2}$. With such high sensitivity, our hybrid nanothermometer provides the capability of measuring millikelvin temperature dynamics with a temporal resolution of a millisecond. For example, Fig. 2d illustrates the histogram of the uncertainty of the measured temperatures with a sampling time of 5 ms. The distribution presents Gaussian statistics with a standard deviation of 1.5 mK. As a comparison, the previous version of nanodiamond-based hybrid sensor (with sensitivity of 11 mK Hz$^{-1/2}$) [30] would need 50 seconds of measurement to achieve the same precision. Nearly two orders of magnitude enhancement of the sensitivity thus enables a wide range of applications, especially in measuring millikelvins temperature change (induced by environment fluctuation, laser heating, and dissipation from micro/nanostructures) with high temporal resolution. To demonstrate the hybrid nanothermometer as a powerful temperature monitor, we performed environment temperature tracking. The environment temperature dynamics at various timescales was measured and shown in Figs. 3a to c. The temperature fluctuation has a maximum amplitude of $\pm 10$ mK, $\pm 5$ mK, and $\pm 2$ mK, at timescales of 100, 1, and 0.1 second, respectively.

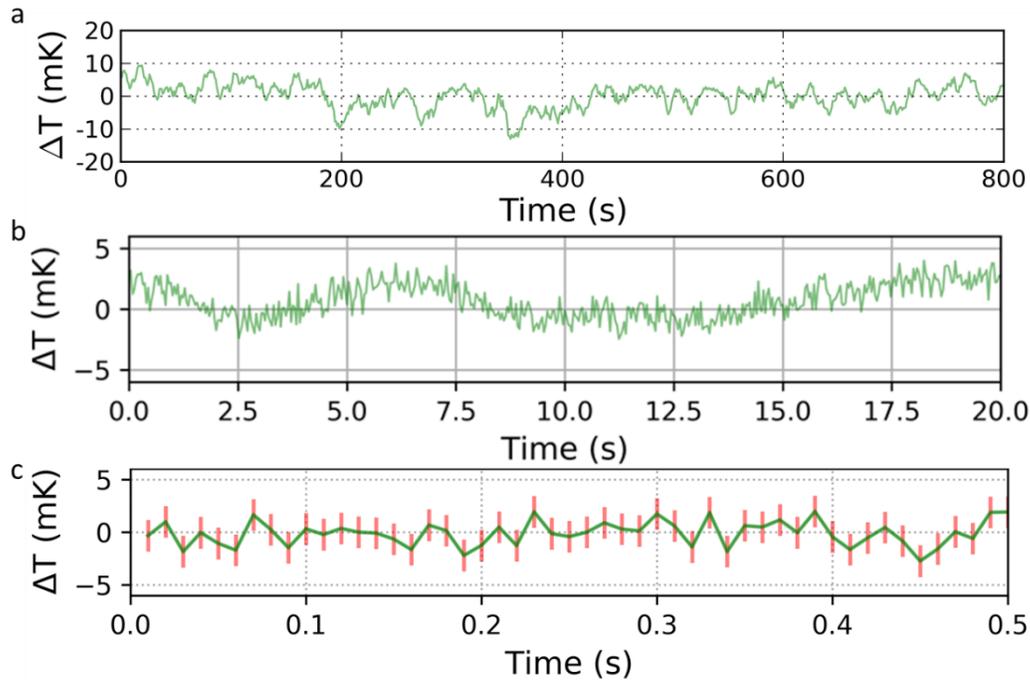

**Figure 3.** Real-time monitoring of local thermal dynamics. **a** to **c**, Environment temperature fluctuation measured by the hybrid nanothermometer with various data integration times (0.5 s, 40 ms and 10 ms, in **a**, **b**, and **c**, respectively).

The hybrid nanothermometer is of potential application in monitoring thermal dynamics in microscopic systems such as biological thermal processes and heat dissipation in micro-/nano-electronic devices. For a proof-of-the-principle experiment, we utilized the microwave antenna around the hybrid nanothermometer as a heating source. We coupled a chopped DC current into the microwave stripline (which has a width of 20 μm and is located ~25 μm away from the sensor) using an RF/DC combiner. The heat generation/dissipation dynamics was monitored by real-time tracking of the local temperature at the location of hybrid nanothermometer. The chopped DC current is illustrated in the upper panel of Fig. 4a and the corresponding ODMR signal are shown in the middle panel. When the DC current is chopped, an instantaneous change in the magnetic field induced by the DC current results in a sudden jump of the spin resonance frequencies in the ODMR, while heating and cooling processes are suggested by the subsequent spin resonance frequency shift after the DC current chopping. The temperature variation is plotted in the lower panel of Fig 4a, in which the $\delta T = 0$ is arbitrarily defined. Figure 4b shows the temperature evolution (averaged over 5 chopping cycles). The temperature increase/decrease of 10 mK with a characteristic time scale ~1 s was clearly observed. In principle, the time scales are determined by several parameters such as the thermal contact between the diamond nanopillar and the antenna, distance between the hybrid sensor and the heating source, and the dissipation rate from the system to the environment. No delay in the heating process was observed, which means that the heat propagation/conduction time from the microwave stripline to the sensor is too short to be resolved in the measurements. The fast heat propagation (with time scale ~1 μs) owes to the high thermal conductivity of the bulk diamond ($> 2200$ W m$^{-1}$ K$^{-1}$) and the short distance ($\approx 30$ μm) between the heater and the sensor. The observation about the thermal dynamics was verified by a control experiment where alternating current with forward and reverse directions were applied with constant heating power. No frequency shift was observed following the jump caused by the electric magnetic fields (see Supplementary Figure 7 for more details). This demonstration experiment proves the potential of the hybrid nanothermometer as a diagnostic tool for studying the thermal dissipation in microelectronics with high spatial and temperature resolution.

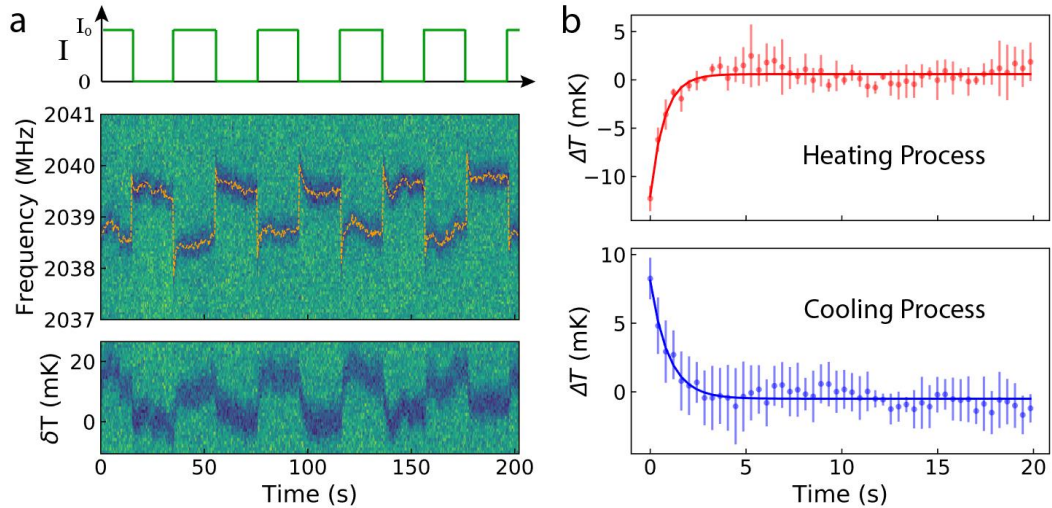

**Figure 4.** Heat dissipation dynamics in the hybrid nanothermometer under pulsed heating. **a**, Upper figure shows the chopped DC current passing through the microwave stripline. Middle figure plots the corresponding ODMR spectra of the NV center in the hybrid nanothermometer. The sudden shift of the ODMR frequency is due to the magnetic field from the chopped DC current. The lower figure is the temperature variation of the hybrid sensor under heating by the chopped current. **b**, The heating and cooling dynamics measured by the hybrid nanothermometer. The $\Delta T = 0$ point is defined by the average of the data at the steady state of the heating/cooling process.

**Conclusion and discussion**

In conclusion, we developed an ultra-sensitive hybrid nanothermometer composed of a single NV center in a diamond nanopillar and a magnetic nanoparticle. When the environment temperature changed near the critical temperature of the MNP, the magnetic field generated by the MNP abruptly changed. The magnetic field change is readily measured by the ODMR of the NV center. The sensitivity of the hybrid nanothermometer is as high as $76~\mu\text{K}~\text{Hz}^{-1/2}$. The high temperature sensitivity indicates fast data acquisition yet with a high temperature measurement precision. We applied the sensor to monitor the environment fluctuations as well as the in-situ heat dissipation dynamics. Stable environment temperature or large dynamic range is critical for further explorations of our hybrid nanothermometer to measure small temperature variation in systems of interest. In fact, the dynamic range of the hybrid sensor can be further enhanced by the frequency-locking scheme for the NV magnetometry [35].

The ultra-sensitive hybrid nanothermometer is especially useful in measuring millikelvin temperature variation with high temporal resolution, which offers a new tool to study a broad range of thermal processes, such as nanoscale chemical reactions and nanoplasmonics, heat dissipation in nano-/micro-electronics, and thermal processes in single cells. The diamond nanopillar can be replaced with a diamond cantilever [36,37] so that a scanning nano-thermometer can be realized with high spatial resolution.

**Methods**

**Experimental setup**
A confocal-AFM correlation microscope was constructed to enable nano-manipulation of single copper-nickel alloy MNPs and in-situ temperature measurements (see Supplementary Figure 1). The AFM scanning head (BioScope Resolve, Bruker) was mounted on the confocal microscope to measure the topography and perform nanomanipulation of the MNP. The ODMR measurements were carried out using a home-built laser scanning confocal microscope. A 532 nm laser was adopted (MGL-III-532-200 mW, CNI) to excite the NV centers. An oil immersion objective lens (Nikon 100x 1.45NA) was used to collect the NV's fluorescence signal, which was then detected by an avalanche photodiode (APD, SPCM-AQRH-15-FC, Excelitas) and counted by a data acquisition (DAQ, PCIe-6363, National Instruments). A microwave (MW) source (N5171B EXG Signal Generator, Keysight) and an amplifier (ZHL-16W-43-S+, Mini-Circuits) were used to generate microwave frequencies for spin measurements. A 20 μm copper wire was used to deliver MW. The sample temperature was controlled by a ceramic heater. The heating area of the heater is about $22 \times 22~\text{mm}^2$ and the pillar was placed in the center with diamond membrane size of $1 \times 1~\text{mm}^2$. Considering the excellent thermal conductivity of diamond material, we assumed the temperature was uniform across the diamond membrane (the distance between reference and hybrid sensor is also below 10 μm). For details see Supplementary Note.

**NV centers in diamond pillars**
In the experiments, high fluorescence intensity of single NV centers in diamond can enhance the sensitivity of the hybrid nanothermometer. Tapered nanopillar shape diamond waveguide was fabricated to achieve enhancement of the fluorescence collection efficiency of single NV centers in diamond. The fabrication process was developed and introduced by S. Momenzadeh et al. [32], where the diamond waveguide was fabricated by electron beam writing and reactive ion etching processes. The optical and spin coherent properties of the NV center are shown in Supplementary Figure 2. For details see Supplementary Note.

**Sensitivity estimation**

FID measurement between the $|m_s = 0\rangle$ and $|m_s = -1\rangle$ state of the NV center in the pillar of the hybrid nanothermometer was applied to estimate the optimal sensitivity as shown in Fig. 2b. The normalized photon count after the FID measurement is [38],

$$S(t) \approx 1 - \frac{C}{2} + \frac{C}{2}\cos(2\pi\delta f t)\exp\left[-\left(\frac{t}{T_2^*}\right)^\nu\right], \qquad (2)$$

where $C$ is the contrast, $\delta f = f - f_p$ is the shifting of the transition frequency $f_-$ from the resonance frequency of the pulses $f_p$, $T_2^*$ is the decoherence time and $\nu$ is the exponent of the decay. Least-square fitting was adopted as shown in Fig. 2b, and the fitting parameters were obtained as $C = 0.27$, $\delta f = 2.7$ MHz, $T_2^* = 1.8$ μs and $\nu = 3.3$. Hence, the shot-noise limited sensitivity of the hybrid nanothermometer was estimated as [38],

$$\eta_T \approx \frac{1}{\sqrt{L_{\text{eff}}}} \left|\frac{df}{dT}\right|^{-1} \left|\frac{dS(t)}{df}\right|^{-1}_{\text{max.}}, \qquad (3)$$

where $L_{\text{eff}} = 9.6 \times 10^4 \text{ s}^{-1}$ is the effective count rate of the measurement and $df/dT = 47$ MHz K$^{-1}$ as shown in inset of Fig. 1d. The optimal waiting time of the measurement is set by maximizing $|dS(t)/df_-|$.

Associated content

Detailed description of the sample fabrication and experiment. This material is available free of charge in the supporting information.

Author information

R.B.L. and Q. L. conceived the idea and supervised the project. C.F.L., K. X., W.H.L., Q.L. and R.B.L. designed the experiments. K.X. and C.F.L. constructed the setup. C.F.L. and K.X. performed the experiments. W.H.L., C.F.L., K.X., Q.L. and R.B.L. analyzed the data. X.F. synthesized the magnetic nanoparticles. A.F., A. D. and J.W. implanted NV centers in diamond and fabricated the diamond nanopillars. C.F. L., W.H.L., K.X., Q.L. and R.B.L. wrote the paper and all authors commented on the paper.

The authors declare no competing financial interest.

Acknowledgements

We would like to thank the discussion with Ning Wang.

Funding

This work was supported by Hong Kong ANR/RGC A-CUHK404/18, and CUHK Group Research Scheme under project code 3110126. AF acknowledges support by the Alexander von Humboldt Foundation. JW acknowledges support by the EU via the ERC grant SMeL, the DFG for support via FOR 2724 and the Volkswagen foundation via the project "Molecular Nanodiamonds"

# Supplementary Information

## of

## Ultra-sensitive hybrid diamond nanothermometer

C.-F. Liu, et. al

## Supplementary Note

**Experimental setup**

We constructed a microscope with correlated functions of confocal microscopy and atomic force microscopy (AFM), to enable nanomanipulation of single copper-nickel alloy magnetic nanoparticles (MNP) and in-situ temperature sensing measurements (see Supplementary Fig. 1). The optically detected magnetic resonance (ODMR) measurements were carried out using a home-built laser scanning confocal microscope. A 532 nm laser was used (MGL-III-532-200 mW, CNI), and a suppression of its power fluctuation down to 0.1% was realized by applying a PID feedback control. An oil immersion objective lens (Nikon 100x 1.45NA) was used to collect the fluorescence of single nitrogen-vacancy (NV) centers in the diamond pillars, which was then detected by an avalanche photodiode (APD, SPCM-AQRH-15-FC, Excelitas) and counted by a data acquisition card (DAQ, PCIe-6363, National Instrument). A pair of APDs were used to measure the second order correlation function of the fluorescence (to check whether the NV centers are good single quantum emitters).

A microwave (MW) source (N5171B EXG Signal Generator, Keysight), a microwave switch (ZASWA_2-50DR+Mini-Circuits), and an amplifier (ZHL-16W-43-S+, Mini-Circuits) were used to generate microwave pulses for spin measurements. A copper wire of width 20 μm was used to deliver the MW. The copper wire, when conducting a DC current, was also used as a heat source in the heat dissipation measurement.

The AFM scanning head (BioScope Resolve, Bruker) was mounted on the confocal microscope to measure the topography and to perform nanomanipulation of MNPs. We first acquired surface topographic images using the AFM tapping mode to localize the positions of the diamond pillars and copper-nickel alloy nanoparticles. The nanomanipulation was achieved by pushing the nanoparticles in the contact mode of the AFM. We re-imaged the surface to check the new positions of the MNPs and acquired the ODMR spectra to evaluate the coupling between the MNPs and the NV centers. Several manipulation processes were applied to optimize the coupling and finally we acquired the surface topography to record the positions of the MNPs and the pillars.

A vibration shielding was built to isolate the correlated microscope from the external environment using sound-absorbing foams. A temperature PID control module was used to keep the temperature fluctuation to < 0.05 °C. The local environment temperature was controlled by a ceramic heater, mounted on top of the glass sample holder.

**NV centers in diamond pillars**

In the experiments, high fluorescence intensity of single NV centers in diamond is required to enhance the sensitivity of the hybrid nanothermometer. Tapered nanopillar shape diamond waveguides were fabricated to improve the fluorescence collection efficiency of single NV centers in diamond. The fabrication process was developed and introduced by S. Momenzadeh et al [1]: the diamond waveguides were fabricated by electron beam writing and reaction ion etching processes. The optical and spin coherent properties of the NV center we studied for the nanothermometer is shown in Supplementary Fig. 2. The second order correlation function of the fluorescence from the pillar $g^{(2)}(\tau)$ is <0.5 at zero delay time (see Supplementary Fig. 2a), which indicates that the studied nanopillar contains only one single NV center. Supplementary Figure 2b plots the saturation curve of the single NV center. Under the excitation of 400 µW power, the saturated fluorescence counts of the single NV center was $1 \text{ M s}^{-1}$. The spin resonance width of the NV center obtained from the ODMR spectra (Supplementary Fig. 2c) is consistent with the dephasing time under free induced decay (FID) measurement, $T_2^* = 1.5$ µs (Supplementary Fig. 2d).

**Temperature calibration**

In the experiment, we calibrated the environment temperature by reference NV centers that were not affected by copper-nickel alloy MNPs (see Supplementary Fig. 3). We first performed the AFM scan to determine that no MNPs were next to the reference NV centers. We hence measured the ODMR spectra. There are four peaks in the ODMR spectra, which correspond to the the $m_s = \pm 1$ electron spin states and the nuclear spin hyperfine coupling. The resonance frequencies were extracted by multi peaks Lorentzian fitting of the ODMR spectra and the zero-field splitting ($D$) was deduced by taking the mean value of the resonance frequencies. Supplementary Fig. 3b shows the ODMR spectra of the reference NV centers for different heating currents of the ceramic heater. We determined the environment temperatures corresponding to the heating current of the ceramic heater using the coefficient as $dD/dT = -74 \text{ kHz K}^{-1}$ (see Supplementary Fig. 3c).

**Reversibility of the nanothermometer in cooling/heating cycles**

We performed heating and cooling processes of the hybrid nanothermometer several times, and recorded the ODMR spectra of the sensor. The spin resonance frequencies of the sensor at different temperature and the corresponding temperature susceptibility for the additional measurements are plotted in Supplementary Fig. 4. Compared with the results in the main text, the sensor shows same spin resonance frequencies during the heating and cooling processes, indicating good reversibility of the hybrid nanothermometer.

**Sensitivity estimation of a few other hybrid nanothermometers**

We have also fabricated two more hybrid nanothermometers composed of single copper-nickel alloy MNPs and single NV centers in diamond nanopillars.

For a nanopillar we detected a single NV center with high fluorescence intensity and long dephasing time, we pushed a single MNP next to the pillar. The AFM images of the hybrid nanothermometers after the AFM nano-manipulation (see insets of Supplementary Figs. 6a and 6d) show that the MNPs were positioned next to the diamond nanopillars. We further measured the spin resonance frequencies of the sensors as functions of the temperature (see Supplementary Figs. 6a and 6d). The temperature susceptibility of each sensor is plotted in Supplementary Figs.

6b and 6e. At the ferromagnetic-paramagnetic transition temperature, the temperature susceptibilities were optimal, being 10 MHz K$^{-1}$ or 7 MHz K$^{-1}$. The NV center spin dephasing times in free induction decay (FID) of the two sensors were measured to be 3.5 μs and 1.6 μs (see Supplementary Figs. 6c and 6f). Using the measured dephasing times 3.5 μs and 1.6 μs, we estimate that these two hybrid nanothermometers have optimal temperature sensitivities of 230 μK Hz$^{-1/2}$ and 550 μK Hz$^{-1/2}$, respectively.

**Temperature sensing study of heat generation and dissipation**

We applied the hybrid nanothermometer to sense the thermal dissipation during heating by a DC current conducted by the copper wire. To confirm that the resonance frequency shifts after the jumping were caused by the temperature dynamics (see Fig. 4 in the manuscript). , we ran the current in both positive and negative directions (the electrical heating effect was maintained the same) and monitored the thermal dissipation by the hybrid nanothermometer (see Supplementary Fig. 7a). In the beginning, we applied the current $I_o$ to the microwave antenna and the current direction was controlled by a microcontroller. When the current through the copper wire switched its direction, a jump of the ODMR spectra was caused by the change of the magnetic field, as seen in Supplementary Fig. 7b. The temperature variation for positive current (Supplementary Fig. 7c) and that for negative current (Supplementary Fig. 7d) present no difference within measurement precision, which confirms that the resonance frequency shift in the manuscript was mainly caused by the temperature variation induced by the pulsed heating current through the copper wire.

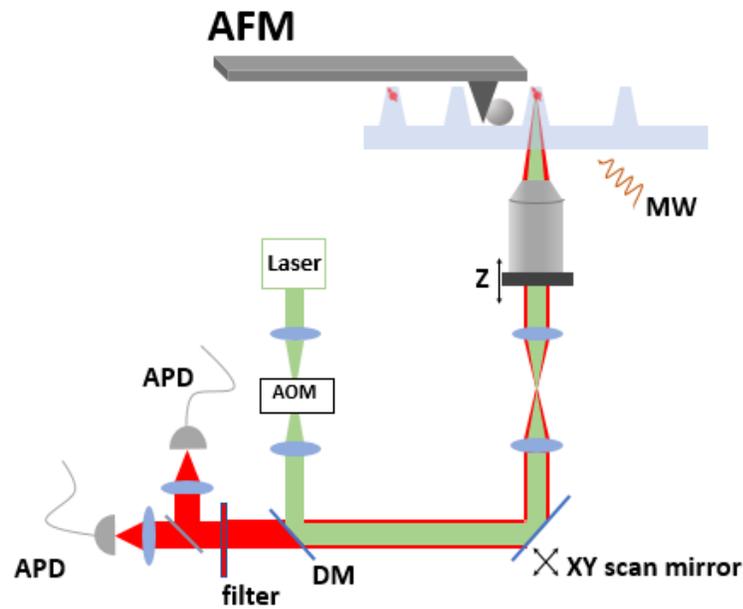

*Supplementary Figure 1. Experimental setup. APD: Avalanche photon diode. AOM: acoustic optical modulate. DM: dichroic mirror. AFM: atomic force microscopy. MW: Microwave.*

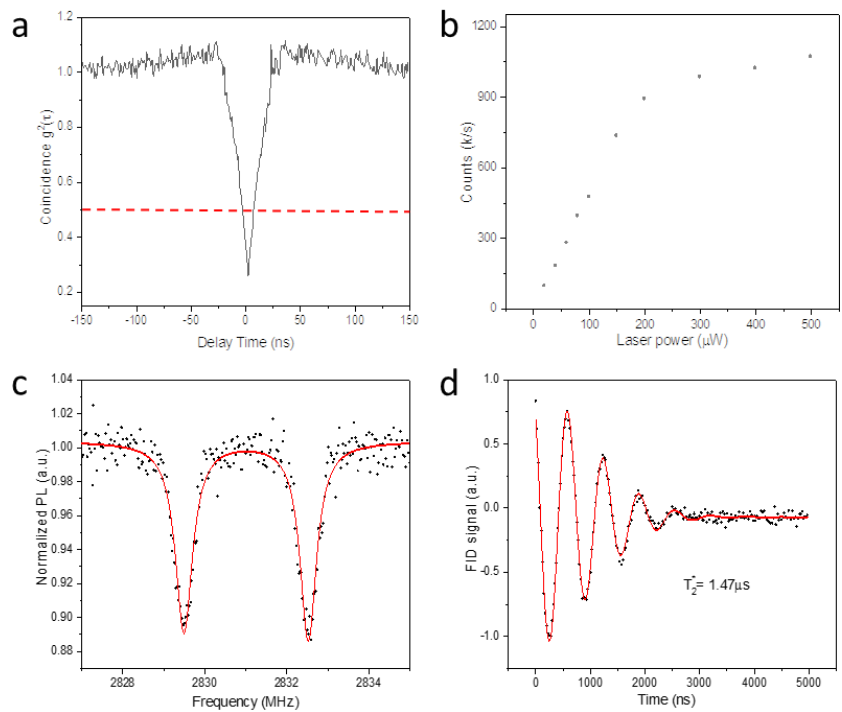

*Supplementary Figure 2. Optical and spin properties of the NV center for the nano-thermometer in main text. (a) Normalized second order correlation function $g^{(2)}(\tau)$ of the fluorescence of the center. (b) Optical saturation behavior of the single NV center. (c) Pulse ODMR spectrum of the NV center under an external magnetic field, showing the $^{15}N$ hyperfine coupling and narrow linewidths. (d) FID of the single NV center with dephasing time $T_2^* = 1.47$ μs.*

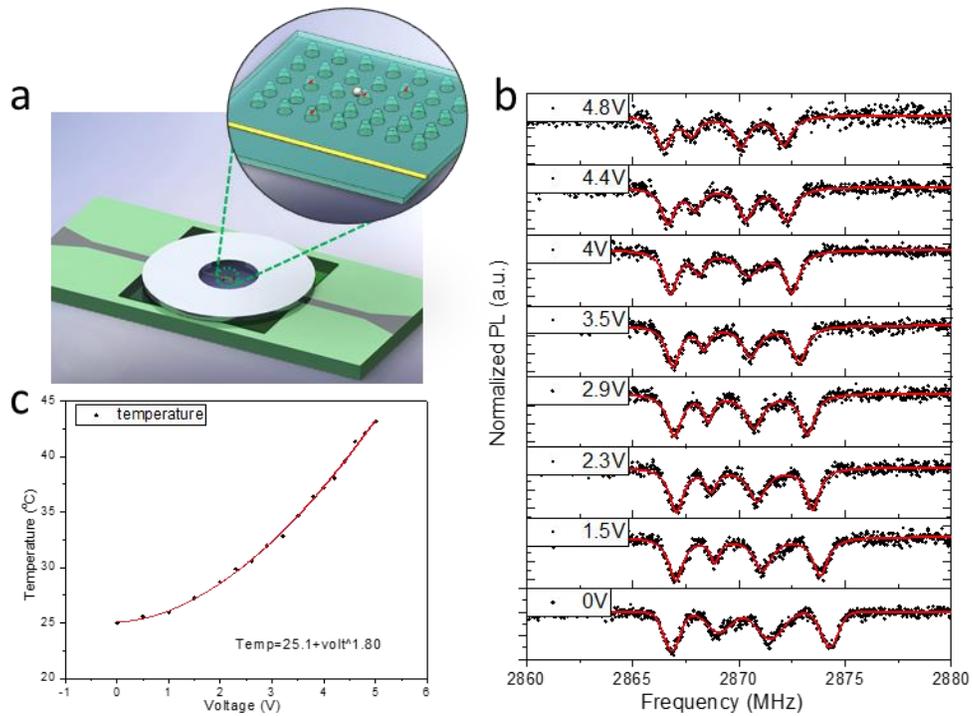

*Supplementary Figure 3. Temperature calibration of the systems. (a) Setup for temperature sensing. The ceramic heater is on top of the diamond sample (white plate). (b) Typical ODMR spectra of a single NV center in the diamond nanopillar under zero external magnetic field (without MNPs in the proximity) for different ceramic heater voltages. (c) Calculated environment temperature as a function of the voltage applied to the ceramic heater.*

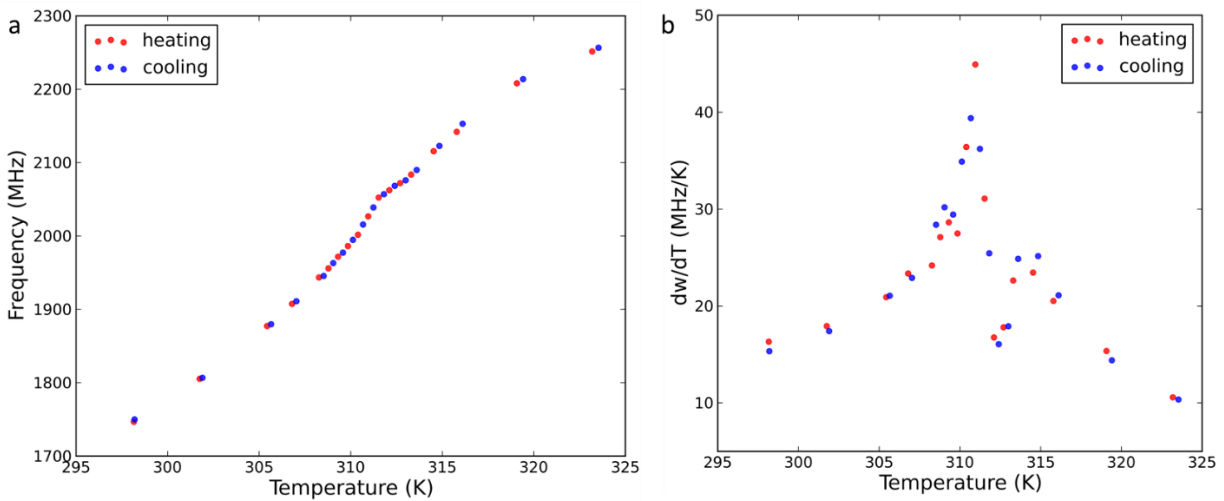

*Supplementary Figure 4. Reversibility of the hybrid nanothermometer. (a) Spin resonance frequencies of the NV center in the hybrid nanothermometer during heating and cooling processes (additional to those present in the main text). (b) Temperature susceptibilities of the hybrid nanothermometer versus the temperature in the additional cooling and heating processes.*

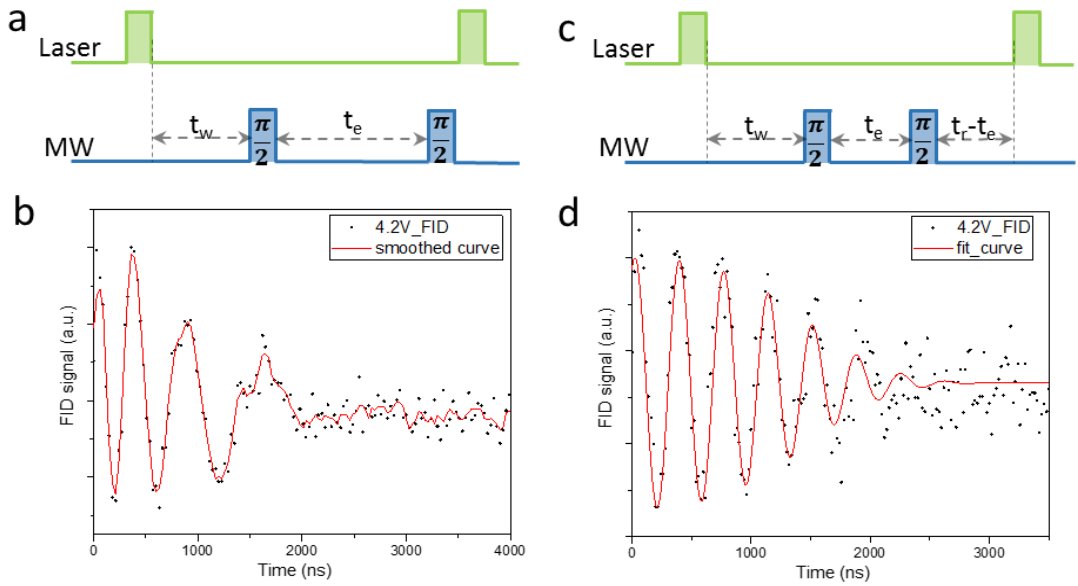

*Supplementary Figure 5. FID signal with and without laser heating duty ratio control. (a) Pulse sequence of the measurement and (b) FID signal without laser heating ratio control. (c) Pulse sequence and (d) FID signal with laser heating control.*

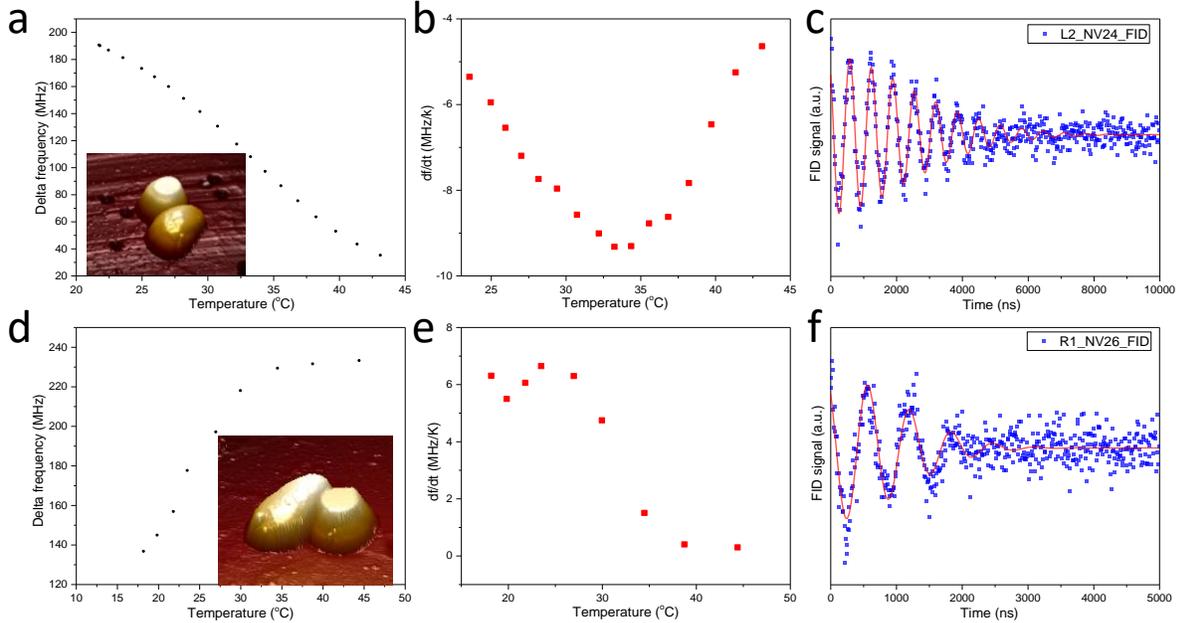

*Supplementary Figure 6. Two additional hybrid sensors. (a) Spin resonance frequencies of the 2$^{nd}$ hybrid nanothermometer versus environment temperature. Inset is the atomic force microscopy (AFM) image of the 2$^{nd}$ hybrid sensor. (b) Temperature susceptibility of the 2$^{nd}$ hybrid nanothermometer versus environment temperature. (c) Free induction decay (FID) of the 2$^{nd}$ hybrid sensor, $T_2^* = 3.5$ μs. (d) Spin resonance frequencies of the 3$^{rd}$ hybrid nanothermometer versus environment temperature. (e) Temperature susceptibility of the 3$^{rd}$ hybrid nanothermometer versus environment temperature. (f) FID of the 3$^{rd}$ hybrid sensor, $T_2^* = 1.6$ μs.*

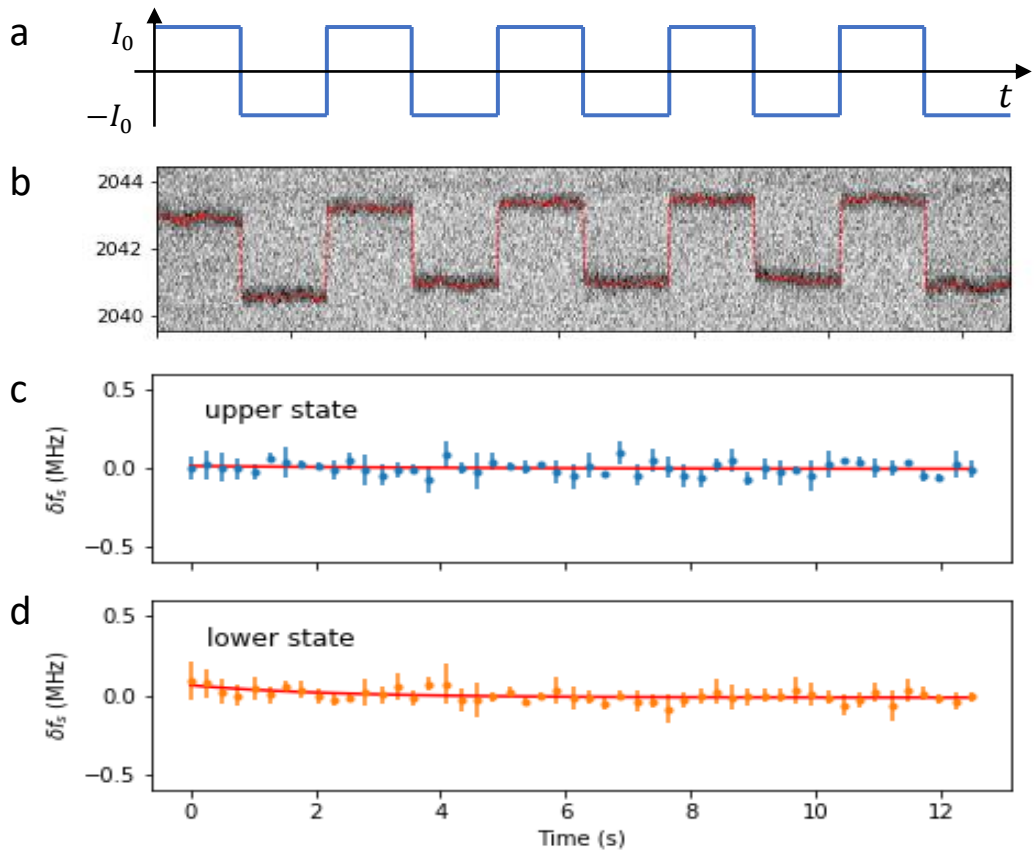

*Supplementary Figure 7. Heating by two opposite currents through the copper wire. (a) Schematic of the current sequence. (b) ODMR spectra of the hybrid nanothermometer when the current was running through the wire. (c) and (d) Sensed resonance frequency variationwhen the current was positive and negative, respectively.*